\documentclass[prl,twocolumn,showpacs,superscriptaddress]{revtex4}

\usepackage{mathrsfs,mathtools,bbold}
\usepackage{amsmath,amssymb}
\usepackage{verbatim}
\usepackage{graphicx}
\usepackage{ccaption}
\usepackage[dvipsnames]{xcolor}
\usepackage[colorlinks=true,linkcolor=blue,citecolor=magenta,urlcolor=MidnightBlue,filecolor=black]{hyperref}

\DeclareFontFamily{OT1}{rsfs}{}
\DeclareFontShape{OT1}{rsfs}{m}{n}{ <-7> rsfs5 <7-10> rsfs7 <10->rsfs10}{} 
\DeclareMathAlphabet{\mycal}{OT1}{rsfs}{m}{n}

\newcommand{\e}{\epsilon}

\newcommand{\C}{\tilde{C}}
\renewcommand{\L}{{\mathcal{L}}}
\newcommand{\bL}{\bar{{\mathcal{L}}}}

\newcommand{\be}[1]{ \begin{equation}\label{#1} }
\newcommand{\ee}{\end{equation}}
\newcommand{\ben}[1]{\begin{eqnarray}\label{#1} }
\newcommand{\een}{\end{eqnarray}}
\newcommand{\p}{\partial}

\newcommand{\w}{\omega}

\newcommand{\refb}[1]{(\ref{#1})}

\renewcommand{\>}{\rangle}

\renewcommand{\a}{\alpha}
\newcommand{\ta}{\tilde{\a}}
\renewcommand{\b}{\beta}
\renewcommand{\t}{\tau}
\newcommand{\s}{\sigma}

\begin{document}
 
\title{The tensionless path from closed to open strings}

\author{Arjun Bagchi}
\email{abagchi@iitk.ac.in}
\affiliation{Indian Institute of Technology, Kanpur 208016, India.}
\affiliation{Blackett Laboratory, Imperial College, London, SW7 2AZ, UK.}

\author{Aritra Banerjee}
\email{aritra@itp.ac.cn}
\affiliation{Institute of Theoretical Physics, Chinese Academy of Sciences,  Beijing 100190, China.}

\author{Pulastya Parekh.}
\email{pulastya@iitk.ac.in}
\affiliation{Indian Institute of Technology, Kanpur 208016, India.}
\affiliation{Erwin Schr$\ddot{o}$dinger International Institute for Mathematics and Physics, 1090 Vienna, Austria.}

\begin{abstract} 
We reconsider the tensionless limit on bosonic closed string theory, where the 3d Bondi-Metzner-Sachs (BMS) algebra appears as symmetries on the worldsheet, as opposed to two copies of the Virasoro algebra in the case of the usual tensile theory. This is an ultra-relativistic limit on the worldsheet. We consider the induced representations of the BMS algebra in the oscillator basis and show that the limit takes the tensile closed string vacuum to the ``induced" vacuum which is identified as a Neumann boundary state. Hence, rather remarkably, an open string emerges from closed strings in the tensionless limit. We also follow the perturbative states in the tensile theory in the limit and show that there is a Bose-Einstein like condensation of all perturbative states on this induced vacuum. This ties up nicely with the picture of the formation of a long string from a gas of strings in the Hagedorn temperature, where the effective string tension goes to zero.    
\end{abstract}


\maketitle

\noindent {\em{\underline{Introduction}}}.  
The very recent first visual evidence of the existence of black holes has reignited interest in the field of gravity and what lies beyond Einstein's theory even in the non-scientific world. Quantum gravity has been the Holy Grail of modern theoretical physics for several decades now. Of the explored avenues, string theory remains the most viable framework to construct a quantum theory of gravity. String theory is endowed with an intrinsic length-scale, the tension of the fundamental string  $T_0 = \frac{1}{2\pi \alpha'}$. In the infinite tension limit, string theory reduces to a quantum field theory and hence loses its stringy-ness. The other extreme limit, the tensionless one, hence explores the ultra-stringy nature of string theory, where the quantum effects of gravity would be the strongest. In this note, this is the regime we are interested in. 

The tensionless regime of string theory has long been the source of great intrigue. This singular limit is the analogue of the massless limit of the point-particle and the strings, as in the point particle case, become null \cite{Schild:1976vq}.  In this limit, it has been long speculated (see e.g. \cite{Francia:2007qt}) that the distinction between closed and open strings become blurred. In this note, we put forward concrete calculations based on the underlying worldsheet symmetries of the tensionless string, to show how, contrary to conventional wisdom, open strings emerge from closed strings in the tensionless limit. We also find a surprising condensation of all perturbative closed string degrees of freedom on the emergent open string, leading us to speculate that this is the indication of a phase transition. 

\medskip

\noindent {\em{\underline{Classical Tensionless Closed Strings}}}. We start with a quick recap of the important features of the classical tensionless closed string theory. The Polyakov action for bosonic tensile string theory is
\be{}
S = -\frac{T}{2}  \int d^2 \xi \sqrt{-g} g^{\alpha \beta} \p_\alpha X^\mu \p_\beta X^\nu \eta_{\mu\nu}.
\ee
The action is invariant under worldsheet diffeomorphisms and gauge fixing is required. It is convenient to fix the conformal gauge $g_{\a \b} = e^{\phi} \eta_{\a \b}$, but there is still some gauge symmetry left over. This residual symmetry is given by two copies of the Virasoro algebra with generators $\L_n, \bL_n$ following:
\be{}
[ \L_m, \L_n] = (m-n) \L_{m+n} + \frac{c}{12} m(m^2 -1) \delta_{m+n,0}. 
\ee
The signature of the tensionless limit is that the worldsheet metric $g^{\alpha \beta}$ degenerates as we take tension to zero. This can be explicitly shown by looking at the Hamiltonian formulation when one equates the phase space action to the Polyakov form \cite{Isberg:1993av}. We implement this by replacing  $T\sqrt{-g} g^{\alpha \beta}$ by $V^\alpha V^\beta$ where $V^\alpha$ is a vector density. The action in the $T \to 0$ limit then becomes \cite{Isberg:1993av}
\be{acti}
S = \int d^2 \xi \,\ V^\alpha V^\beta \p_\alpha X^\mu \p_\beta X^\nu \eta_{\mu\nu}.
\ee
This action is again invariant under world-sheet diffeomorphisms and one needs to fix gauge. In the tensionless analogue of the conformal gauge $V^\alpha = (1, 0)$, there is again a residual symmetry, which in this case turns out to be the 3d Bondi-Metzner-Sachs algebra (or equivalently the 2d Galilean Conformal Algebra) \cite{Isberg:1993av, Gamboa:1989zc, Bagchi:2013bga}:
\ben{bms3}
&& [L_m, L_n] = (m-n) L_{m+n} + \frac{c_L}{12} m(m^2 -1) \delta_{m+n,0},  \nonumber\\
&& [L_m, M_n] = (m-n) M_{m+n} + \frac{c_M}{12} m(m^2 -1) \delta_{m+n,0}, \nonumber\\
&& [M_m, M_n] = 0.
\een
This algebra also appears as the asymptotic symmetries of 3d flat spacetimes at null infinity \cite{Barnich:2006av} (thus has been used in studies of holography in flat spacetimes, see e.g. \cite{Bagchi:2010eg, Bagchi:2012xr}) and also in non-relativistic conformal systems \cite{Bagchi:2009my, Bagchi:2009pe}. In eq \refb{bms3} above, $c_L, c_M$ are central charges consistent with Jacobi identities. 

In the tensionless limit, the fundamental string grows long and floppy, and the length of the string becomes infinite. For the co-ordinates on the world-sheet, this limit is best viewed as \cite{Bagchi:2013bga} 
\be{urlim} 
\t \to \e \t, \ \s \to \s, \,\ \mbox{with} \,\ \e\to 0.  
\ee  
This is an ultra-relativistic (UR) or Carrollian limit on the worldsheet \cite{Bagchi:2013bga, Duval:2014lpa}, where the (worldsheet) speed of light goes to zero. It can also be thought of as an infinite boost that makes an ordinary string into a null string. In terms of the generators, this takes the form: 
\be{ultra-lim}
L_n =\L_n - \bL_{-n}, \quad M_n =\e (\L_n + \bL_{-n}).
\ee
We now turn to mode expansions of the bosonic tensionless string. In the $V^\a=(1,0)$ gauge, the action can be used to derive equations of motion and constraints:  
\be{eom}
{\p_\t}^2 X^\mu = 0; \quad ({\p_\t {X}})^2 = \p_\t X \cdot \p_\s X = 0.
\ee
The EOM can be solved to yield the mode expansion \cite{Bagchi:2015nca}: 
\be{mode}
X^{\mu}(\sigma,\tau)=x^{\mu}+\sqrt{2c'}B^{\mu}_0\tau
+\sqrt{2c'}\sum_{n\neq0}\frac{i}{n}\left(A^{\mu}_n-in\tau B^{\mu}_n\right)e^{-in\sigma} 
\ee
Here we have already put in closed string boundary conditions $X^\mu(\sigma,\tau) = X^\mu(\sigma + 2\pi,\tau)$. Defining 
\be{lmab}
L_n=\sum_{m} A_{-m}\cdot B_{m+n},\quad M_n=\sum_{m} B_{-m}\cdot B_{m+n}
\ee
the constraint equations in \refb{eom} become
\begin{subequations}
\ben{} 
T_{(1)}&=&\sum_{n} \left(L_n-in\tau M_n \right) \ e^{-in\sigma}=0, \\ 
T_{(2)}&=& \sum_{n} M_n\ e^{-in\sigma}=0.
\een
\end{subequations}
These combinations are the energy-momentum tensors of a 2d BMS invariant field theory and can be derived just from the symmetry algebra. The classical algebra of the oscillator modes $(A,B)$ is: 
\be{}
\{A^{\mu}_m,A^{\nu}_n\}=\{B^{\mu}_m,B^{\nu}_n\}=0, \{A^{\mu}_m,B^{\nu}_n\}=-2im\delta_{m+n} \eta^{\mu \nu}. 
\ee
This is not the algebra of harmonic oscillator modes. This will be an important point as we go forward. One can obtain the mode expansion \refb{mode} by looking at the tensile mode expansion and using the limit \refb{urlim}. This relates the tensile oscillators $(\a, \ta)$ to $(A, B)$:
\be{aal}
A^{\mu}_n=\frac{1}{\sqrt{\e}}({\alpha}^{\mu}_n-\tilde{\alpha}^{\mu}_{-n}), \quad B^{\mu}_n=\sqrt{\e}({\alpha}^{\mu}_n+\tilde{\alpha}^{\mu}_{-n}). 
\ee 
It can be easily seen that using \refb{lmab} and the above relation \refb{aal}, we get back \refb{ultra-lim}. 

\medskip

\noindent {\em{\underline{BMS Representation Theory}}}. We now turn to the representation theory of the symmetry that underlines the theory of tensionless strings, viz. the BMS$_3$ algebra. Relevant work in this direction includes \cite{Barnich:2014kra, Barnich:2015uva, Campoleoni:2016vsh}. An important class of representations for the BMS$_3$ algebra is the so-called {\em{massive modules}} \cite{Campoleoni:2016vsh}. The Hilbert space of such a module contains a wavefunction $|M,s\>$ satisfying:
\begin{subequations}\label{ind}
\ben{}
&& M_0 |M,s\> = M |M,s\>, \, L_0 |M,s\> = s |M,s\>; \\ 
&& M_n |M,s\> = 0, \ \forall n \neq 0.
\een
\end{subequations}
This defines a 1d representation of the subalgebra of BMS$_3$ spanned by $\{L_0, M_n, c_L, c_M\}$. This can be used to define an {\em{induced BMS module}} with basis vectors 
\be{}
|\Psi\>= L_{n_1} L_{n_2} \ldots L_{n_k} |M,s\>.
\ee 
Here $n_1\geq n_2 \geq \ldots \geq n_k$ are integers which can be both positive or negative. We wish to now understand how this is related to the parent 2d CFT representations. In the highest weight representations of the Virasoro algebra, primary states $|h, \bar{h} \>$ are given by: 
\begin{subequations}
\ben{}
&& \L_0 |h, \bar{h} \> = h |h, \bar{h} \>, \ \bL_0 |h, \bar{h} \> = \bar{h}  |h, \bar{h} \>; \\
&& \L_n |h, \bar{h} \> = 0 = \bL_n |h, \bar{h} \>, \ n>0. \, 
\een 
\end{subequations}
Following the UR limit \refb{ultra-lim}, this translates to 
\be{}
\left(L_n + \frac{1}{\e} M_n \right)|h, \bar{h} \> = \left(L_{-n} - \frac{1}{\e} M_{-n} \right) |h, \bar{h} \> =0, \ n>0. 
\ee
In the limit $\e\to0$, this gives \refb{ind}, along with the identification: 
\be{}
M = \e (h + \bar{h}), \ s= h - \bar{h}.
\ee
So highest weight representations of 2d CFTs become induced representations of BMS invariant field theories in the UR limit. In terms of the oscillator modes, the induced modules are defined by 
\be{} 
B_n |M, s\> = 0, \, \forall n \neq 0, \quad B_0^2 |M, s\> = M |M, s\>. 
\ee

\medskip

\noindent {\em{\underline{Tensionless strings and induced representations}}}. We had remarked that the $(A,B)$ oscillators were not in a harmonic oscillator basis. To rectify this, we define: 
\be{} 
C^{\mu}_{\ n} = \frac{1}{2}({A}^{\mu}_n+B^{\mu}_{n}), \quad \C^{\mu}_{\ n}=\frac{1}{2}(-{A}^{\mu}_{-n}+B^{\mu}_{-n}) 
\ee
The algebra of the modes now becomes that of two decoupled harmonic oscillators:
$$[C^{\mu}_m,C^{\nu}_n]=m\delta_{m+n,0}\eta^{\mu\nu}, \  [\C^{\mu}_m,\C^{\nu}_n]=m\delta_{m+n,0}\eta^{\mu\nu}. $$
The tensile and tensionless raising and lowering operators are related by
\ben{}
&& C^{\mu}_n(\e) =\beta_+ {\alpha}^{\mu}_n+\beta_- \tilde{\alpha}^{\mu}_{- n} \nonumber \\
&&\C^{\mu}_n(\epsilon) =\beta_- {\alpha}^{\mu}_{-n}+\beta_+\tilde{\alpha}^{\mu}_{ n}. 
\een
where 
\be{}
\beta_\pm=\frac{1}{2}\left(\sqrt{\epsilon}\pm\frac{1}{\sqrt{\epsilon}}\right)
\ee
It is clear that since there is a mixing of tensile raising and lowering operators in $C, \C$, the $C$ vacuum $|0\>_c$ defined by 
\be{}
|0\>_c: C^{\mu}_{\ n} |0\>_c = 0 = \C^{\mu}_{\ n} |0\>_c \quad \forall n>0.
\ee
is different from tensile vacuum $|0\>_\a$ which in turn is defined by
\be{}
|0\>_\a: \a^{\mu}_{\ n} |0\>_\a = 0 = \ta^{\mu}_{\ n} |0\>_\a \quad \forall n>0.
\ee
Let us now turn our attention to the vacuum in the induced representation, which we denote by $|I\>$. We have
\be{indB}
B_n^\mu |I \> = 0, \ \forall n. 
\ee
Since there is no ordering ambiguity in $M_0$ when acting on this vacuum, the mass of the induced vacuum has to be zero. In terms of $C$ oscillators, the induced vacuum conditions are:
\be{}
(C^{\mu}_n + \C^{\mu}_{\ -n}) |I\> = 0, \quad \forall n. 
\ee
This is the condition of a {\em{Neumann boundary state}} and the solution is given by 
\be{indvac}
|I\> = \mathcal{N} \prod_{n=1}^{\infty} \exp \left( - \frac{1}{n} C_{-n} \cdot \C_{-n} \right) |0\>_c 
\ee
where $ \mathcal{N}$ is a (infinite) normalisation constant.

\medskip

\noindent {\em{\underline{From closed to open strings}}}. 
The relation between the $C$-oscillators and the $\a$-oscillators is a Bogoliubov transformation on the worldsheet: 
\ben{}
&& \a^{\mu}_n = e^{i G} C_{n} e^{-iG} =\cosh\theta \ {C}^{\mu}_n -\sinh\theta \ \tilde{C}^{\mu}_{- n}, \\
&& \tilde{\a}^{\mu}_n= e^{i G} \tilde{C}_{n} e^{-iG}=-\sinh\theta \ {C}^{\mu}_{-n}+\cosh\theta \ \tilde{C}^{\mu}_{ n}, \nonumber
\een
where 
\be{}
G = i \sum_{n=1}^{\infty} \theta \left[ C_{-n}.\tilde{C}_{-n} -C_n .\tilde{C}_n\right], \ \tanh \theta = \frac{\e-1}{\e+1}. 
\ee
We can use this to relate the two vacua: 
\ben{}
|0\>_\a &=&  \exp[i G] |0\>_c  \\
&=&  \left(\frac{1}{\cosh\theta}\right)^{1+1+\hdots} \prod_{n=1}^{\infty}  \exp[\frac{\tanh\theta}{n} C_{-n}\tilde{C}_{-n}]  |0\>_c \nonumber
\een
Using the regularisation: $1+1+1+\hdots \infty=\zeta(0)=-\frac{1}{2}$, we finally get
 \be{sqz}
 |0\>_\a= \sqrt{\cosh\theta} \prod_{n=1}^{\infty}  \exp\left[\frac{\tanh\theta}{n} \, C_{-n} \cdot \tilde{C}_{-n}\right]  |0\>_c
 \ee
From the point of view of $|0\>_c$, $|0\>_\a$ is a squeezed state.

\begin{figure}[t]
{\includegraphics[scale=0.22]{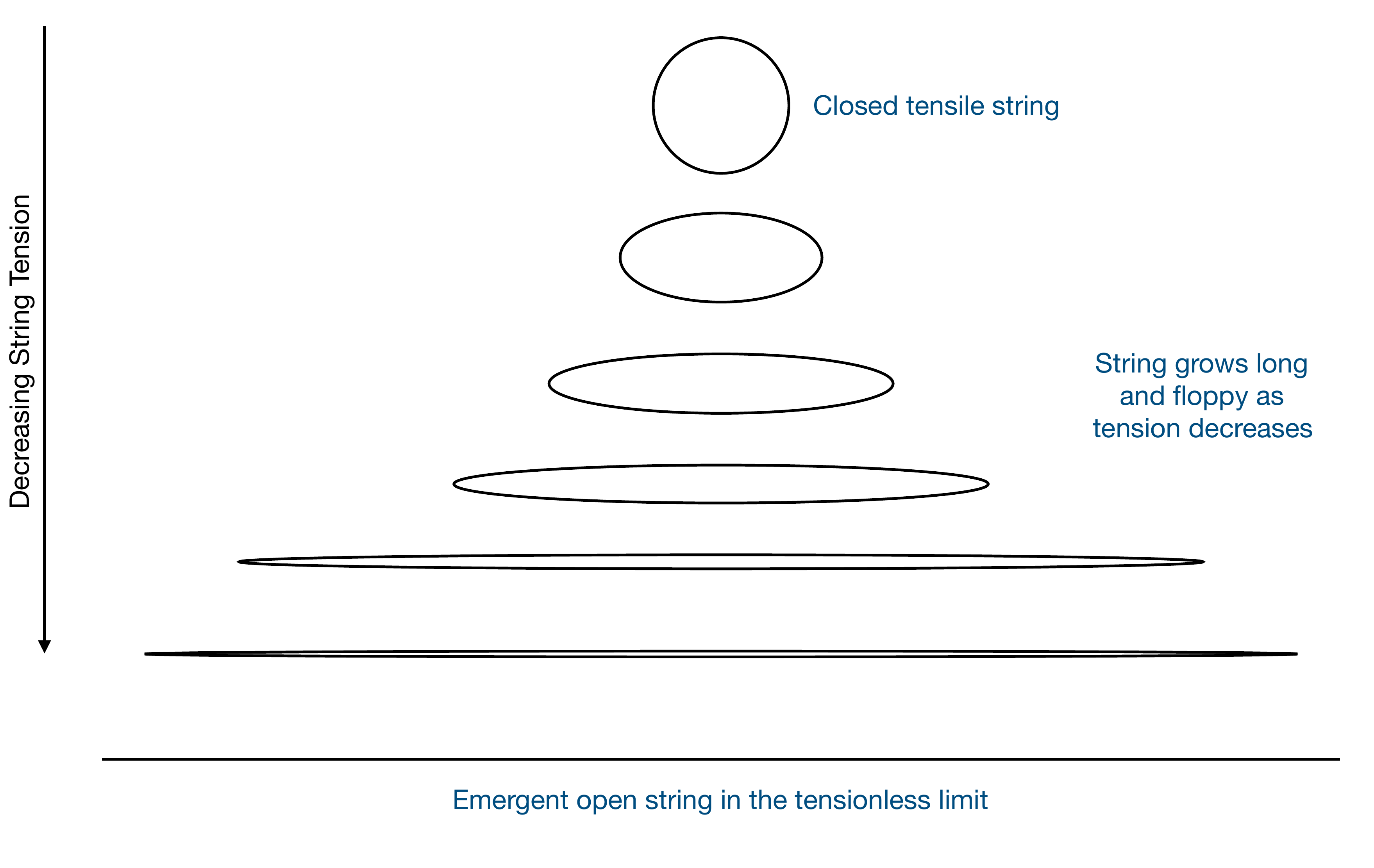}}
\caption{Formation of open strings from closed strings.}\label{fig1}
\end{figure}

Now let us elucidate how an open string emerges from a closed string as we take tension to zero. When $\e=1$, $\tanh \theta =0$, and we have $|0\>_\a = |0\>_c$. This is the closed string vacuum. As $\e$ changes from 1, from the point of view of the $C$ observer, the vacuum evolves. It becomes a squeezed state as shown in \refb{sqz}. In the limit where $\e \to 0$, we have $\tanh \theta = -1$. The relation is thus:
\be{}
|0\>_\a= {\mathcal{N}} \prod_{n=1}^{\infty}  \exp\left[- \, \frac{1}{n} C_{-n} \cdot \tilde{C}_{-n}\right]  |0\>_c
\ee
This is exactly the {\em{induced vacuum $|I\>$}} that we introduced in \refb{indB}--\refb{indvac}.  As we mentioned there, this is a Neumann boundary state. This is thus an {\em{open string}} free to move in all dimensions {\footnote{See e.g. \cite{Blumenhagen:2009zz} for a discussion of boundary states in CFT in the context of string theory.}}. We have thus obtained {\em{an open string}} by taking {\em{a tensionless limit}} on a {\em{closed string theory}}. Physically, how this is happening can be visualised as in Figure \refb{fig1}. 

An open string with Neumann boundary conditions in all directions can also be interpreted as a space-filling D-25 brane. An intuitive picture of how a closed string becomes a spacefilling brane is shown below in Fig. \refb{fig2}.

\begin{figure}[h]
{\includegraphics[scale=0.25]{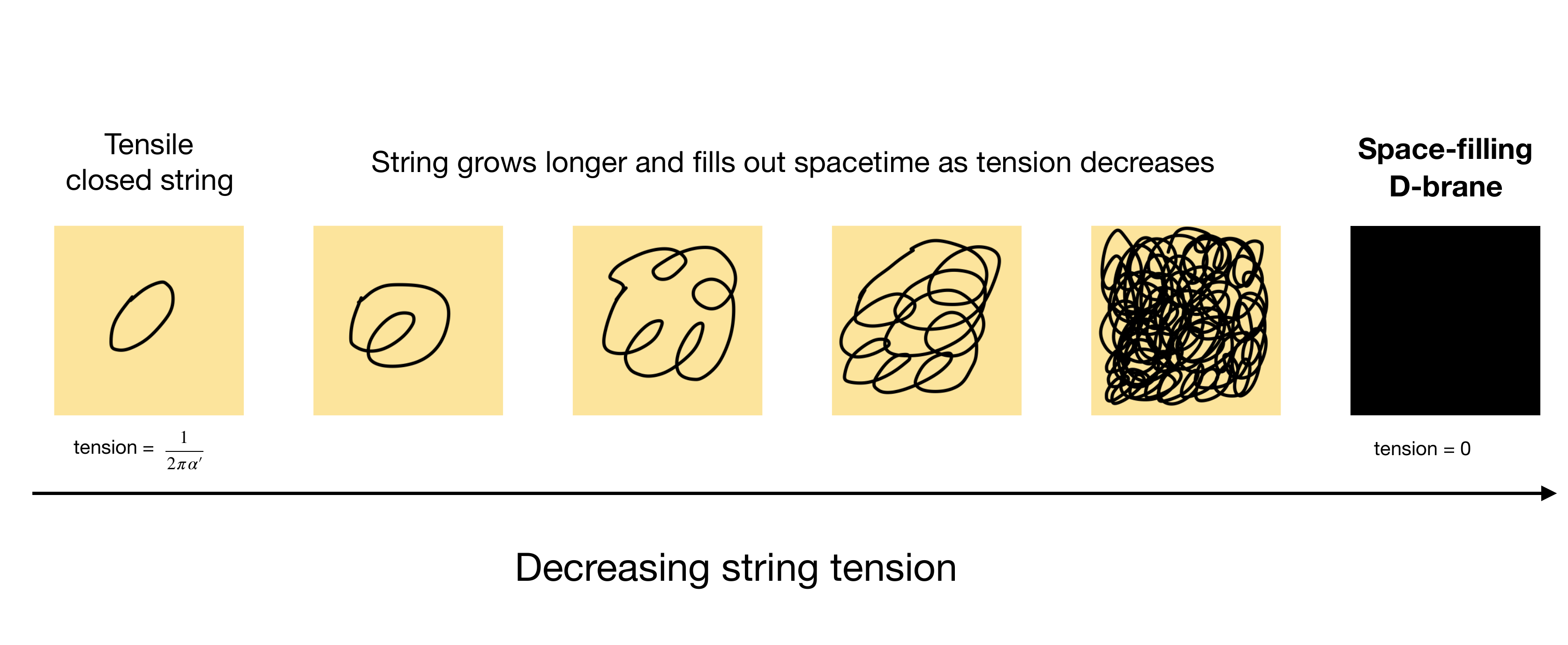}}
\caption{Formation of spacefilling D-brane from closed strings.}\label{fig2}
\end{figure}

Computing the central charges of the residual symmetry algebra in this induced vacuum $|I\>$, we are led to conclude that $c_L=c_M=0$. For $c_M=0$, we can use an analysis of null vectors in the BMS$_3$ algebra to show that there is a truncation of the algebra to its Virasoro sub-algebra \cite{Bagchi:2009pe}. Thus in this case, when we dial $\e$ away from 1, the symmetry algebra stays two copies of the Virasoro algebra, until it reaches $\e=0$, where it becomes BMS$_3$, which in turn reduces to a single copy of the Virasoro due to the absence of the central term $c_M$. So even from this perspective, there is a clear hint of an open string appearing from the closed string worldsheet as the tension is dialled down to zero.

\medskip

\noindent {\em{\underline{Bose-Einstein Condensation on the Worldsheet}}}. We now describe a novel process by which this emergent open string is formed from the states of the tensile closed string theory. Consider any perturbative state in the original tensile theory $|\Psi\> = \xi_{\mu\nu} \a^\mu_{-n} \tilde{\a}^\nu_{-n} |0\>_\a$ where $\xi_{\mu\nu}$ is a polarisation tensor. Let us attempt to understand the evolution of the state as $\e\to0$. 
Close to $\e=0$, the alpha vacuum can be approximated as follows: $$ |0\>_\a = |I\> + \e |I_1\> + \e^2 |I_2\> + \ldots .$$ In this limit, the conditions on the alpha vacuum $(\a_n |0\>_\a =  \tilde{\a}_n |0\>_\a = 0, \ n>0)$ translate to: 
\ben{cond}
&& B_n |I\> = 0, \quad \forall n\neq 0; \\ 
&& A_n |I\> + B_n |I_1\> = 0, \ A_{-n} |I\> - B_{-n} |I_1\> =0, \ n>0.  \nonumber
\een
One can now take this limit on the state $|\Psi\> = \a_{-n} \tilde{\a}_{-n} |0\>_\a$ (where now we have suppressed the spacetime indices) which is now rewritten as
\be{}
|\Psi\>  = \frac{1}{\e}\left(B_{-n} + {\e} A_{-n}\right) \left(B_{n} - \e A_{n}\right) (|I\> + \e |I_1\> + \ldots). \nonumber
\ee
Using the commutation relations and \refb{cond}, we can show
\be{}
|\Psi\> \to K |I\> \quad \mbox{as} \ \e\to0, 
\ee
where $K$ is a level dependent constant: $K=2 n \eta^{\mu\nu}\xi_{\mu\nu}$.  Thus, {\em{all perturbative closed string states condense on the open string induced vacuum}}. This condensation is like a Bose-Einstein condensation on the worldsheet and is indicative of a phase transition. A point to note here is that this is independent of the level of the state and hence very high energy  perturbative states also condense to this vacuum. 

\begin{figure}
{\includegraphics[scale=0.22]{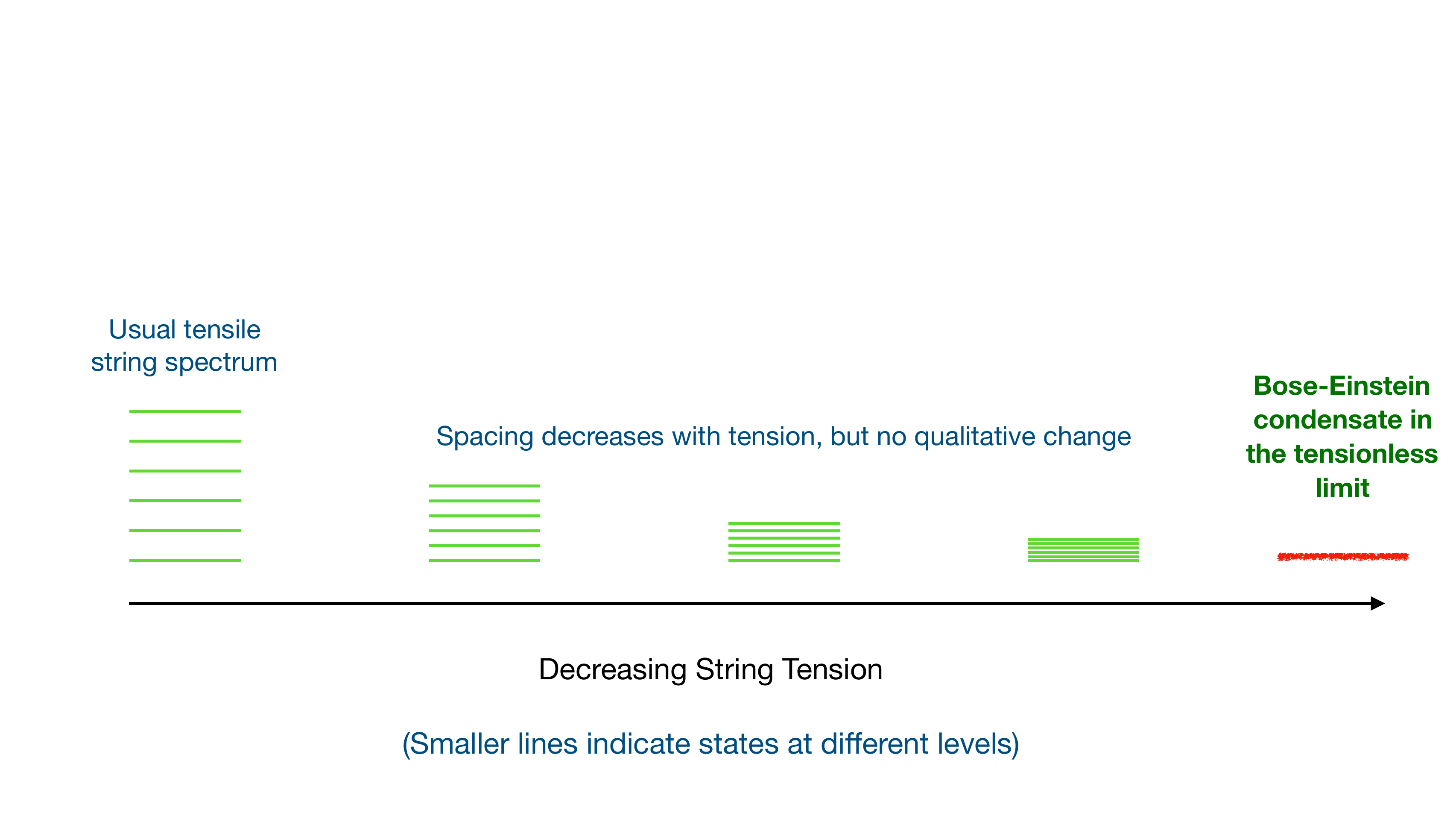}}
\caption{BE condensation on the worldsheet.}
\end{figure}

\medskip

\noindent {\em{\underline{Connections with Hagedorn Physics}}}.
Let us remind the reader that the framework of tensionless strings is useful for addressing questions of string theory near the Hagedorn temperature $\mathcal{T}_H$. The Hagedorn temperature is the point where the partition function of the single particle states in string theory blows up and it has been long speculated that this is indicative of a phase transition to a new phase where very different degrees of freedom arise \cite{Atick:1988si}. When looking at the theory of free strings, the Hagedorn phase transition can be understood as follows. This is the regime where it become thermodynamically favourable to form a long string as opposed to heating up a gas of strings \cite{Giddings:1989xe, Lowe:1994nm}. Strings near the Hagedorn temperature become effectively tensionless \cite{Pisarski:1982cn, Olesen:1985ej}:
\be{}
T_{\mbox{\tiny{eff}}} = T_0 \sqrt{1 - \frac{\mathcal{T}^2}{\mathcal{T}_H^2}}
\ee
Here $T_{\mbox{\tiny{eff}}}$ is the effective tension, $T_0 = 1/2\pi\a'$ is the usual tension of the string, and $\mathcal{T}$ is the temperature of the system. 
We propose that the {\em{induced vacuum $|I\>$ is the emergent long string from the point of view of the worldsheet}}. 

The Bose-Einstein condensation on the worldsheet described above is also the worldsheet manifestation of the Hagedorn phase transition. This seems to be at odds with the observation that the BE condensation is something that happens at absolute zero while the Hagedorn phase transition is a very high energy phenomenon. To clarify this, we remind the reader that the Hagedorn temperature is related to the string tension: 
$\mathcal{T}_H = \frac{1}{2\pi \sqrt{2\a'}}$. So the tensionless limit, which is $\a'\to \infty$, from the point of view of the worldsheet, drives the Hagedorn temperature to zero and hence relates to the above described BE condensation.

\medskip

\noindent {\em{\underline{Summary and Future Directions}}}. We have shown the rather remarkable emergence of an open string from closed strings in the tensionless limit in the context of bosonic string theory and also shown that there is a condensation of all perturbative closed string modes to form this open string. It would be of interest to take this beyond bosonic string theory and generalise to the case of superstrings. From the point of view of the worldsheet, there are two different classical manifestations of the tensionless superstring, which arise from two different contractions of the fermionic generators, which have been dubbed the homogeneous \cite{Lindstrom:1990ar, Bagchi:2016yyf} and the inhomogeneous tensionless superstring \cite{Bagchi:2017cte}. We wish to examine both these limits and study the analogue of the induced representations of the underlying super BMS algebras and see what emerges in the quantum regime. 

The classical aspects of the tensionless strings, we believe, is now well understood and the central feature is the emergence of the BMS algebra on the null worldsheet. The Riemannian structure of the tensile worldsheet is replaced by an emergent Carrollian structure. This is similar to what happens for other null surfaces, like the null boundary of flatspace \cite{Hartong:2015xda, Hartong:2015usd, Bagchi:2019xfx} and also black hole horizons \cite{Donnay:2019jiz}. But the quantum mechanical structure is much more intricate. We have, in this paper, shed light on one of the possible vacua, the induced vacuum. Depending on the choice of vacuum structure, the resulting quantum theory would be very different. A detailed exposition of this would be presented in a companion paper \cite{toappear}. There we would also elaborate on other aspects of the induced vacuum.

We have just skimmed the surface of the representation theoretic aspects of the underlying BMS algebra for the tensionless string. There has been a recent resurgence of flat space physics relating asymptotic symmetries with soft theorems and memory effects (see e.g. \cite{Strominger:2017zoo} for a review). The story in three spacetime dimensions has not yet been properly fleshed out {\footnote{See however \cite{Hijano:2019qmi} for an initiation of work in this direction.}}, and this is in part because of the lack of physical degrees of freedom of gravity in bulk dimensions $d=3$. The interplay between BMS symmetries and soft theorems in $d=3$ would have very interesting consequences for the study of tensionless strings. The analogues of soft theorems on the worldsheet may tie in with the infinite number of relations between string amplitudes in the very high energy or equivalently the tensionless limit of strings \cite{Gross:1987kza, Gross:1987ar, Gross:1988ue}. These remain active lines of inquiry and we hope to shed light on these aspects in the future. 

\medskip

\noindent {\em{\underline{Acknowledgements}}}.
We would like to thank Shankhadeep Chakrabortty for a wonderful ongoing collaboration on aspects of tensionless strings and discussions in this particular project as well. Discussions with R. Basu, A. Campoleoni, R. Gopakumar, J. Hartong, N. Iqbal, and J. Simon are also gratefully acknowledged. 

AB thanks Durham University and Imperial College London for hosting him during the course of this work. This work was presented at the ESI workshop ``Higher spins and Holography" in Vienna, the University of Amsterdam, and the University of Geneva. AB thanks the respective groups for discussions and warm hospitality. PP thanks Institute for Theoretical Physics, TU Wien for hospitality. AB was supported by a Senior CO-FUND fellowship of the Durham Institute of Advanced Study, and a MATRICS grant from SERB, India. ArB is supported in part by the Chinese Academy of Sciences (CAS) Hundred-Talent Program, by the Key Research Program of Frontier Sciences, CAS, and by Project 11647601 supported by NSFC. PP is funded by the Junior Research Fellowship Programme from ESI Vienna.

\bibliography{ref}

\begin{thebibliography}{34}
\expandafter\ifx\csname natexlab\endcsname\relax\def\natexlab#1{#1}\fi
\expandafter\ifx\csname bibnamefont\endcsname\relax
  \def\bibnamefont#1{#1}\fi
\expandafter\ifx\csname bibfnamefont\endcsname\relax
  \def\bibfnamefont#1{#1}\fi
\expandafter\ifx\csname citenamefont\endcsname\relax
  \def\citenamefont#1{#1}\fi
\expandafter\ifx\csname url\endcsname\relax
  \def\url#1{\texttt{#1}}\fi
\expandafter\ifx\csname urlprefix\endcsname\relax\def\urlprefix{URL }\fi
\providecommand{\bibinfo}[2]{#2}
\providecommand{\eprint}[2][]{\url{#2}}

\bibitem[{\citenamefont{Schild}(1977)}]{Schild:1976vq}
\bibinfo{author}{\bibfnamefont{A.}~\bibnamefont{Schild}},
  \bibinfo{journal}{Phys. Rev.} \textbf{\bibinfo{volume}{D16}},
  \bibinfo{pages}{1722} (\bibinfo{year}{1977}).

\bibitem[{\citenamefont{Francia et~al.}(2007)\citenamefont{Francia, Mourad, and
  Sagnotti}}]{Francia:2007qt}
\bibinfo{author}{\bibfnamefont{D.}~\bibnamefont{Francia}},
  \bibinfo{author}{\bibfnamefont{J.}~\bibnamefont{Mourad}}, \bibnamefont{and}
  \bibinfo{author}{\bibfnamefont{A.}~\bibnamefont{Sagnotti}},
  \bibinfo{journal}{Nucl. Phys.} \textbf{\bibinfo{volume}{B773}},
  \bibinfo{pages}{203} (\bibinfo{year}{2007}), \eprint{hep-th/0701163}.

\bibitem[{\citenamefont{Isberg et~al.}(1994)\citenamefont{Isberg, Lindstrom,
  Sundborg, and Theodoridis}}]{Isberg:1993av}
\bibinfo{author}{\bibfnamefont{J.}~\bibnamefont{Isberg}},
  \bibinfo{author}{\bibfnamefont{U.}~\bibnamefont{Lindstrom}},
  \bibinfo{author}{\bibfnamefont{B.}~\bibnamefont{Sundborg}}, \bibnamefont{and}
  \bibinfo{author}{\bibfnamefont{G.}~\bibnamefont{Theodoridis}},
  \bibinfo{journal}{Nucl. Phys.} \textbf{\bibinfo{volume}{B411}},
  \bibinfo{pages}{122} (\bibinfo{year}{1994}), \eprint{hep-th/9307108}.

\bibitem[{\citenamefont{Gamboa et~al.}(1990)\citenamefont{Gamboa, Ramirez, and
  Ruiz-Altaba}}]{Gamboa:1989zc}
\bibinfo{author}{\bibfnamefont{J.}~\bibnamefont{Gamboa}},
  \bibinfo{author}{\bibfnamefont{C.}~\bibnamefont{Ramirez}}, \bibnamefont{and}
  \bibinfo{author}{\bibfnamefont{M.}~\bibnamefont{Ruiz-Altaba}},
  \bibinfo{journal}{Nucl. Phys.} \textbf{\bibinfo{volume}{B338}},
  \bibinfo{pages}{143} (\bibinfo{year}{1990}).

\bibitem[{\citenamefont{Bagchi}(2013)}]{Bagchi:2013bga}
\bibinfo{author}{\bibfnamefont{A.}~\bibnamefont{Bagchi}},
  \bibinfo{journal}{JHEP} \textbf{\bibinfo{volume}{05}}, \bibinfo{pages}{141}
  (\bibinfo{year}{2013}), \eprint{1303.0291}.

\bibitem[{\citenamefont{Barnich and Compere}(2007)}]{Barnich:2006av}
\bibinfo{author}{\bibfnamefont{G.}~\bibnamefont{Barnich}} \bibnamefont{and}
  \bibinfo{author}{\bibfnamefont{G.}~\bibnamefont{Compere}},
  \bibinfo{journal}{Class. Quant. Grav.} \textbf{\bibinfo{volume}{24}},
  \bibinfo{pages}{F15} (\bibinfo{year}{2007}), \eprint{gr-qc/0610130}.

\bibitem[{\citenamefont{Bagchi}(2010)}]{Bagchi:2010eg}
\bibinfo{author}{\bibfnamefont{A.}~\bibnamefont{Bagchi}},
  \bibinfo{journal}{Phys. Rev. Lett.} \textbf{\bibinfo{volume}{105}},
  \bibinfo{pages}{171601} (\bibinfo{year}{2010}), \eprint{1006.3354}.

\bibitem[{\citenamefont{Bagchi et~al.}(2013)\citenamefont{Bagchi, Detournay,
  Fareghbal, and Sim{\'o}n}}]{Bagchi:2012xr}
\bibinfo{author}{\bibfnamefont{A.}~\bibnamefont{Bagchi}},
  \bibinfo{author}{\bibfnamefont{S.}~\bibnamefont{Detournay}},
  \bibinfo{author}{\bibfnamefont{R.}~\bibnamefont{Fareghbal}},
  \bibnamefont{and}
  \bibinfo{author}{\bibfnamefont{J.}~\bibnamefont{Sim{\'o}n}},
  \bibinfo{journal}{Phys. Rev. Lett.} \textbf{\bibinfo{volume}{110}},
  \bibinfo{pages}{141302} (\bibinfo{year}{2013}), \eprint{1208.4372}.

\bibitem[{\citenamefont{Bagchi and Gopakumar}(2009)}]{Bagchi:2009my}
\bibinfo{author}{\bibfnamefont{A.}~\bibnamefont{Bagchi}} \bibnamefont{and}
  \bibinfo{author}{\bibfnamefont{R.}~\bibnamefont{Gopakumar}},
  \bibinfo{journal}{JHEP} \textbf{\bibinfo{volume}{07}}, \bibinfo{pages}{037}
  (\bibinfo{year}{2009}), \eprint{0902.1385}.

\bibitem[{\citenamefont{Bagchi et~al.}(2010)\citenamefont{Bagchi, Gopakumar,
  Mandal, and Miwa}}]{Bagchi:2009pe}
\bibinfo{author}{\bibfnamefont{A.}~\bibnamefont{Bagchi}},
  \bibinfo{author}{\bibfnamefont{R.}~\bibnamefont{Gopakumar}},
  \bibinfo{author}{\bibfnamefont{I.}~\bibnamefont{Mandal}}, \bibnamefont{and}
  \bibinfo{author}{\bibfnamefont{A.}~\bibnamefont{Miwa}},
  \bibinfo{journal}{JHEP} \textbf{\bibinfo{volume}{08}}, \bibinfo{pages}{004}
  (\bibinfo{year}{2010}), \eprint{0912.1090}.

\bibitem[{\citenamefont{Duval et~al.}(2014)\citenamefont{Duval, Gibbons, and
  Horvathy}}]{Duval:2014lpa}
\bibinfo{author}{\bibfnamefont{C.}~\bibnamefont{Duval}},
  \bibinfo{author}{\bibfnamefont{G.~W.} \bibnamefont{Gibbons}},
  \bibnamefont{and} \bibinfo{author}{\bibfnamefont{P.~A.}
  \bibnamefont{Horvathy}}, \bibinfo{journal}{J. Phys.}
  \textbf{\bibinfo{volume}{A47}}, \bibinfo{pages}{335204}
  (\bibinfo{year}{2014}), \eprint{1403.4213}.

\bibitem[{\citenamefont{Bagchi et~al.}(2016{\natexlab{a}})\citenamefont{Bagchi,
  Chakrabortty, and Parekh}}]{Bagchi:2015nca}
\bibinfo{author}{\bibfnamefont{A.}~\bibnamefont{Bagchi}},
  \bibinfo{author}{\bibfnamefont{S.}~\bibnamefont{Chakrabortty}},
  \bibnamefont{and} \bibinfo{author}{\bibfnamefont{P.}~\bibnamefont{Parekh}},
  \bibinfo{journal}{JHEP} \textbf{\bibinfo{volume}{01}}, \bibinfo{pages}{158}
  (\bibinfo{year}{2016}{\natexlab{a}}), \eprint{1507.04361}.

\bibitem[{\citenamefont{Barnich and Oblak}(2014)}]{Barnich:2014kra}
\bibinfo{author}{\bibfnamefont{G.}~\bibnamefont{Barnich}} \bibnamefont{and}
  \bibinfo{author}{\bibfnamefont{B.}~\bibnamefont{Oblak}},
  \bibinfo{journal}{JHEP} \textbf{\bibinfo{volume}{06}}, \bibinfo{pages}{129}
  (\bibinfo{year}{2014}), \eprint{1403.5803}.

\bibitem[{\citenamefont{Barnich and Oblak}(2015)}]{Barnich:2015uva}
\bibinfo{author}{\bibfnamefont{G.}~\bibnamefont{Barnich}} \bibnamefont{and}
  \bibinfo{author}{\bibfnamefont{B.}~\bibnamefont{Oblak}},
  \bibinfo{journal}{JHEP} \textbf{\bibinfo{volume}{03}}, \bibinfo{pages}{033}
  (\bibinfo{year}{2015}), \eprint{1502.00010}.

\bibitem[{\citenamefont{Campoleoni et~al.}(2016)\citenamefont{Campoleoni,
  Gonzalez, Oblak, and Riegler}}]{Campoleoni:2016vsh}
\bibinfo{author}{\bibfnamefont{A.}~\bibnamefont{Campoleoni}},
  \bibinfo{author}{\bibfnamefont{H.~A.} \bibnamefont{Gonzalez}},
  \bibinfo{author}{\bibfnamefont{B.}~\bibnamefont{Oblak}}, \bibnamefont{and}
  \bibinfo{author}{\bibfnamefont{M.}~\bibnamefont{Riegler}},
  \bibinfo{journal}{Int. J. Mod. Phys.} \textbf{\bibinfo{volume}{A31}},
  \bibinfo{pages}{1650068} (\bibinfo{year}{2016}), \eprint{1603.03812}.

\bibitem[{\citenamefont{Atick and Witten}(1988)}]{Atick:1988si}
\bibinfo{author}{\bibfnamefont{J.~J.} \bibnamefont{Atick}} \bibnamefont{and}
  \bibinfo{author}{\bibfnamefont{E.}~\bibnamefont{Witten}},
  \bibinfo{journal}{Nucl. Phys.} \textbf{\bibinfo{volume}{B310}},
  \bibinfo{pages}{291} (\bibinfo{year}{1988}).

\bibitem[{\citenamefont{Giddings}(1989)}]{Giddings:1989xe}
\bibinfo{author}{\bibfnamefont{S.~B.} \bibnamefont{Giddings}},
  \bibinfo{journal}{Phys. Lett.} \textbf{\bibinfo{volume}{B226}},
  \bibinfo{pages}{55} (\bibinfo{year}{1989}).

\bibitem[{\citenamefont{Lowe and Thorlacius}(1995)}]{Lowe:1994nm}
\bibinfo{author}{\bibfnamefont{D.~A.} \bibnamefont{Lowe}} \bibnamefont{and}
  \bibinfo{author}{\bibfnamefont{L.}~\bibnamefont{Thorlacius}},
  \bibinfo{journal}{Phys. Rev.} \textbf{\bibinfo{volume}{D51}},
  \bibinfo{pages}{665} (\bibinfo{year}{1995}), \eprint{hep-th/9408134}.

\bibitem[{\citenamefont{Pisarski and Alvarez}(1982)}]{Pisarski:1982cn}
\bibinfo{author}{\bibfnamefont{R.~D.} \bibnamefont{Pisarski}} \bibnamefont{and}
  \bibinfo{author}{\bibfnamefont{O.}~\bibnamefont{Alvarez}},
  \bibinfo{journal}{Phys. Rev.} \textbf{\bibinfo{volume}{D26}},
  \bibinfo{pages}{3735} (\bibinfo{year}{1982}).

\bibitem[{\citenamefont{Olesen}(1985)}]{Olesen:1985ej}
\bibinfo{author}{\bibfnamefont{P.}~\bibnamefont{Olesen}},
  \bibinfo{journal}{Phys. Lett.} \textbf{\bibinfo{volume}{160B}},
  \bibinfo{pages}{408} (\bibinfo{year}{1985}).

\bibitem[{\citenamefont{Lindstrom et~al.}(1991)\citenamefont{Lindstrom,
  Sundborg, and Theodoridis}}]{Lindstrom:1990ar}
\bibinfo{author}{\bibfnamefont{U.}~\bibnamefont{Lindstrom}},
  \bibinfo{author}{\bibfnamefont{B.}~\bibnamefont{Sundborg}}, \bibnamefont{and}
  \bibinfo{author}{\bibfnamefont{G.}~\bibnamefont{Theodoridis}},
  \bibinfo{journal}{Phys. Lett.} \textbf{\bibinfo{volume}{B258}},
  \bibinfo{pages}{331} (\bibinfo{year}{1991}).

\bibitem[{\citenamefont{Bagchi et~al.}(2016{\natexlab{b}})\citenamefont{Bagchi,
  Chakrabortty, and Parekh}}]{Bagchi:2016yyf}
\bibinfo{author}{\bibfnamefont{A.}~\bibnamefont{Bagchi}},
  \bibinfo{author}{\bibfnamefont{S.}~\bibnamefont{Chakrabortty}},
  \bibnamefont{and} \bibinfo{author}{\bibfnamefont{P.}~\bibnamefont{Parekh}},
  \bibinfo{journal}{JHEP} \textbf{\bibinfo{volume}{10}}, \bibinfo{pages}{113}
  (\bibinfo{year}{2016}{\natexlab{b}}), \eprint{1606.09628}.

\bibitem[{\citenamefont{Bagchi et~al.}(2018)\citenamefont{Bagchi, Banerjee,
  Chakrabortty, and Parekh}}]{Bagchi:2017cte}
\bibinfo{author}{\bibfnamefont{A.}~\bibnamefont{Bagchi}},
  \bibinfo{author}{\bibfnamefont{A.}~\bibnamefont{Banerjee}},
  \bibinfo{author}{\bibfnamefont{S.}~\bibnamefont{Chakrabortty}},
  \bibnamefont{and} \bibinfo{author}{\bibfnamefont{P.}~\bibnamefont{Parekh}},
  \bibinfo{journal}{JHEP} \textbf{\bibinfo{volume}{02}}, \bibinfo{pages}{065}
  (\bibinfo{year}{2018}), \eprint{1710.03482}.

\bibitem[{\citenamefont{Hartong}(2015)}]{Hartong:2015xda}
\bibinfo{author}{\bibfnamefont{J.}~\bibnamefont{Hartong}},
  \bibinfo{journal}{JHEP} \textbf{\bibinfo{volume}{08}}, \bibinfo{pages}{069}
  (\bibinfo{year}{2015}), \eprint{1505.05011}.

\bibitem[{\citenamefont{Hartong}(2016)}]{Hartong:2015usd}
\bibinfo{author}{\bibfnamefont{J.}~\bibnamefont{Hartong}},
  \bibinfo{journal}{JHEP} \textbf{\bibinfo{volume}{10}}, \bibinfo{pages}{104}
  (\bibinfo{year}{2016}), \eprint{1511.01387}.

\bibitem[{\citenamefont{Bagchi et~al.}(2019)\citenamefont{Bagchi, Mehra, and
  Nandi}}]{Bagchi:2019xfx}
\bibinfo{author}{\bibfnamefont{A.}~\bibnamefont{Bagchi}},
  \bibinfo{author}{\bibfnamefont{A.}~\bibnamefont{Mehra}}, \bibnamefont{and}
  \bibinfo{author}{\bibfnamefont{P.}~\bibnamefont{Nandi}}
  (\bibinfo{year}{2019}), \eprint{1901.10147}.

\bibitem[{\citenamefont{Donnay and Marteau}(2019)}]{Donnay:2019jiz}
\bibinfo{author}{\bibfnamefont{L.}~\bibnamefont{Donnay}} \bibnamefont{and}
  \bibinfo{author}{\bibfnamefont{C.}~\bibnamefont{Marteau}}
  (\bibinfo{year}{2019}), \eprint{1903.09654}.

\bibitem[{\citenamefont{Bagchi et~al.}(To appear)\citenamefont{Bagchi,
  Banerjee, Chakrabortty, Dutta, and Parekh}}]{toappear}
\bibinfo{author}{\bibfnamefont{A.}~\bibnamefont{Bagchi}},
  \bibinfo{author}{\bibfnamefont{A.}~\bibnamefont{Banerjee}},
  \bibinfo{author}{\bibfnamefont{S.}~\bibnamefont{Chakrabortty}},
  \bibinfo{author}{\bibfnamefont{S.}~\bibnamefont{Dutta}}, \bibnamefont{and}
  \bibinfo{author}{\bibfnamefont{P.}~\bibnamefont{Parekh}} (\bibinfo{year}{To
  appear}).

\bibitem[{\citenamefont{Strominger}(2017)}]{Strominger:2017zoo}
\bibinfo{author}{\bibfnamefont{A.}~\bibnamefont{Strominger}}
  (\bibinfo{year}{2017}), \eprint{1703.05448}.

\bibitem[{\citenamefont{Gross and Mende}(1987)}]{Gross:1987kza}
\bibinfo{author}{\bibfnamefont{D.~J.} \bibnamefont{Gross}} \bibnamefont{and}
  \bibinfo{author}{\bibfnamefont{P.~F.} \bibnamefont{Mende}},
  \bibinfo{journal}{Phys. Lett.} \textbf{\bibinfo{volume}{B197}},
  \bibinfo{pages}{129} (\bibinfo{year}{1987}).

\bibitem[{\citenamefont{Gross and Mende}(1988)}]{Gross:1987ar}
\bibinfo{author}{\bibfnamefont{D.~J.} \bibnamefont{Gross}} \bibnamefont{and}
  \bibinfo{author}{\bibfnamefont{P.~F.} \bibnamefont{Mende}},
  \bibinfo{journal}{Nucl. Phys.} \textbf{\bibinfo{volume}{B303}},
  \bibinfo{pages}{407} (\bibinfo{year}{1988}).

\bibitem[{\citenamefont{Gross}(1988)}]{Gross:1988ue}
\bibinfo{author}{\bibfnamefont{D.~J.} \bibnamefont{Gross}},
  \bibinfo{journal}{Phys. Rev. Lett.} \textbf{\bibinfo{volume}{60}},
  \bibinfo{pages}{1229} (\bibinfo{year}{1988}).

\bibitem[{\citenamefont{Blumenhagen and Plauschinn}(2009)}]{Blumenhagen:2009zz}
\bibinfo{author}{\bibfnamefont{R.}~\bibnamefont{Blumenhagen}} \bibnamefont{and}
  \bibinfo{author}{\bibfnamefont{E.}~\bibnamefont{Plauschinn}},
  \bibinfo{journal}{Lect. Notes Phys.} \textbf{\bibinfo{volume}{779}},
  \bibinfo{pages}{1} (\bibinfo{year}{2009}).

\bibitem[{\citenamefont{Hijano}(2019)}]{Hijano:2019qmi}
\bibinfo{author}{\bibfnamefont{E.}~\bibnamefont{Hijano}}
  (\bibinfo{year}{2019}), \eprint{1905.02729}.

\end{thebibliography}


\end{document}